\documentstyle [psfig,epsf,graphicx]{mn}
\newcommand{\mincir}{\raise
-3.truept\hbox{\rlap{\hbox{$\sim$}}\raise4.truept\hbox{$<$}\ }}
\newcommand{\magcir}{\raise
-3.truept\hbox{\rlap{\hbox{$\sim$}}\raise4.truept\hbox{$>$}\ }}
\newcommand{\minmag}{\raise
-3.truept\hbox{\rlap{\hbox{$<$}}\raise5.truept\hbox{$<$}\ }}
\newcommand{\be}{\begin{equation}}
\newcommand{\ee}{\end{equation}}
 \newcommand{\ba}{\begin{eqnarray}}
\newcommand{\ea}{\end{eqnarray}}
\newcommand{\brr}{\begin{array}}
\newcommand{\err}{\end{array}}
\newcommand{\bc}{\begin{center}}
\newcommand{\ec}{\end{center}}

\title[The XMM-Newton/2dF  extended source sample]
{The XMM-Newton/2dF survey-VIII: The extended X-ray sources}
\author[T. Gaga et al.]{T. Gaga$^{1,2}$, M. Plionis$^{1,3}$, 
 S. Basilakos$^1$,  I. Georgantopoulos$^{1}$, A. Georgakakis$^{1}$
\\
\vspace{0.1cm}
  $^1$ Institute of Astronomy \& Astrophysics, National Observatory
  of Athens, I. Metaxa \& V. Pavlou, Athens, 15236, Greece \\
  $^2$ Section of Astrophysics, Astronomy and Mechanics,
  Department of Physics, University of Athens, Panepistimiopolis, \\
  GR-157 83, Zografos, Athens, Greece\\
  $^3$ Instituto Nacional de Astrofisica, Optica y Electronica (INAOE) 
Apartado Postal 51 y 216, 72000, Puebla, Pue., Mexico\\
} 

\begin{document}
\maketitle

\begin{abstract}
We present a sample of eight extended X-ray sources detected in
the wide-field  ($\sim 2.3 {\rm deg}^2$), bright (2-10 ksec)
{\it XMM-Newton}/2dF survey, reaching  a flux limit of 
 $\sim 2\times10^{-14}$ erg s$^{-1}$ cm$^{-2}$.
Of these, seven are identified as secure X-ray clusters 
in the soft 0.3-2 keV band using a standard wavelet algorithm
on either the PN or the MOS images.
Spectroscopic or photometric redshifts are
available for five clusters, spanning a range between 0.12 and 0.68. The X-ray 
spectral fittings show temperatures between 1 and 4.6 keV,
characteristic of poor clusters and groups of galaxies. 
We derive for the first time the {\it XMM-Newton} cluster 
 number count $\log N-\log S$ distribution 
albeit with poor statistics. Both the  $\log N-\log S$
 and the Luminosity-Temperature relation 
  are  in  good agreement with previous {\it ROSAT} results.
\end{abstract}


\section{Introduction}\label{sec_intro}
Surveys of cluster of galaxies have been constructed using a variety of
identification algorithms, applied mostly on optical
wide-field observations (eg. Abell, Corwin \& Olowin 1989; 
Lumsden et al. 1992; Dalton et al. 1994; Postman et al. 1996;
Olsen et al. 1999; Gladders \& Yee 2000;
Goto et al. 2002; Bahcall et al. 2003).
Although the recent cluster identification algorithms are increasingly 
more sophisticated, the optical surveys are known to suffer from
projection effects and related biases. 

Alternatively, X-ray selected cluster samples have a number of
advantages with respect to surveys based in the optical. 
The detection of the diffuse Intra-Cluster Medium (ICM), which 
emits strongly at X-ray wavelengths, can be securely associated with
clusters of galaxies, since the X-ray emission 
traces the central part of the cluster
and is proportional to the square of the hot gas density. This fact results
in a high contrast with respect to the X-ray
background. 

A number of recent works have attempted to understand
cluster selection procedures applied on different wavelength
data and quantify their differences
(eg. Donahue et al 2002; Basilakos et al 2004 and references
therein). In a recent paper we have investigated the X-ray properties
of a subset of the Cut and Enhance (CE) SDSS clusters of Goto et
al. (2002) using seven public {\it XMM-Newton} pointings (Plionis et al. 2005). We
have found that only 8 out of the 17 CE clusters, in the areas
investigated, appear in X-rays
with $f_x \magcir 1.2 \times 10^{-14}$ ergs cm$^{-2}$ s$^{-1}$.

The great advantage of X-ray selected clusters is that they
are less susceptible to projection effects while due to their 
strong ICM emission they can also be
observed to very large distances. They have a well defined
flux-limit from which it is relatively straight forward to derive
their redshift selection function which is instrumental in measuring
their clustering properties and the evolution of their physical and
dynamical state. These studies can then provide us with a wealth of
cosmological information (eg. Borgani \& Guzzo 2000; Rosati,  Borgani
\& Norman 2002) which in conjunction with the WMAP results should
provide a consistent cosmological framework.
 
The {\it Einstein} as well as the {\it ROSAT} 
satellites, supplemented by follow-up studies with 
{\it Asca} and {\it Beppo-Sax} 
allowed a leap forward in the cluster X-ray astronomy, producing large samples
of both nearby and distant clusters (eg. Stocke et al. 1991;
Castander et al. 1995; Ebeling 
et al. 1996a, 1996b; Scharf et al. 1997; de Grandi et al. 1999; 
Ebeling et al. 2000; B\"{o}hringer et al. 2001; Gioia et al. 2001; 
Rosati, Borgani \& Norman 2002; Moretti et al. 2004). 
The last few years  
the {\it XMM-Newton} and the {\it Chandra} observatories, 
with much larger effective area and better spatial 
resolution than the previous missions are playing a key role in the 
detection of relatively distant clusters. For example, the 
medium-deep  Large Scale Survey  is aimed in studying 
the evolution of the cluster-cluster
correlation function and the cluster number density out to $z\sim 1$
(Pierre et al. 2004; Valtchanov et al. 2004). A serendipitous survey, 
conducted by Kolokotronis et al. (2005) using public {\it XMM-Newton} data and
wide-field optical multiband follow-ups, aims to investigate
not only the global properties of the X-ray cluster distribution but also
the optical and X-ray properties of individual 
clusters at different redshifts (see also Land et al. 2005 for an
{\it XMM-Newton}-SDSS cluster survey). 
Note that with integrations of a few tens of ksec, the {\it XMM-Newton}
can reach to a sensitivity of $\sim 5 \times 10^{-15}$ erg cm$^{-2}$
s$^{-1}$ and detect rich clusters ($L_x \sim 10^{44} \; h^{-2}$ erg $s^{-1}$)
as extended sources out to $z\sim 2$ (eg. Pierre et al. 2004).

In this work we use the {\it XMM-Newton}/2dF survey 
(see Georgakakis et al. 2004) which covers an area 
of $\rm \simeq 2.3\,deg^{2}$ and reaches a flux limit of 
$2\times 10^{-14}$ erg cm$^{-2}$ s$^{-1}$.   
 This flux depth is comparable to the RDCS {\it ROSAT} survey of 
 Rosati et al. (1998). 
The main aim of this work is to find candidate clusters 
from their extended emission, producing an X-ray selected cluster catalogue.
In Section \ref{sec_survey} we present our {\it XMM-Newton}/2dF survey,
we describe the reduction of the X-ray data and the cluster
identification method.
In section \ref{final} we present the cluster flux and
temperature estimation method, the details of the candidate 
clusters, the corresponding $\log N- \log S$ and
luminosity-temperature relations.
Finally in section \ref{conclusions} we summarise our conclusions. 
Throughout this paper we adopt $H_{\circ}=100  h$ km s$^{-1}$ Mpc$^{-1}$,
 $\Omega_{\rm m}=0.3$ and $\Omega_{\Lambda}=0.7$ and
note that all X-ray luminosities and fluxes are 
reported in the 0.3-2 keV energy band.

\section{The XMM-Newton/2dF Survey}\label{sec_survey}
Two regions near the  North and South Galactic Pole regions 
(F864; RA(J2000)=$13^{\rm h}41^{\rm m}$; Dec.(J2000)=$00\degr00\arcmin$
 and SGP; RA(J2000)=$00^{\rm h} 57^{\rm m}$, Dec.(J2000)=$-28\degr
00\arcmin$, respectively) 
were surveyed by the  {\it XMM-Newton} between May 2002
and February 2003 as part of the guaranteed time program. 
Details of the observations, data reduction procedures 
 and overall project aims can be found
in Georgakakis et al. (2004). Briefly, we note that the observations
consist of a total of 18 pointings equally split between the SGP 
and the F864 regions each  with an exposure time
between $\approx2-10$\,ks.
Our survey overlaps with the 2dF Galaxy Redshift
Survey (2dFGRS\footnote{http://msowww.anu.edu.au/2dFGRS/}; Colless
et al. 2001) and hence, high quality spectra, redshifts and spectral
 classifications are available for all $bj<19.4$\,mag galaxies
 in our regions. 
Note that the F864 region overlaps with the Sloan Digital Sky Survey (York
et al. 2000), 
which provides optical photometry in 5 bands ($ugriz$;  Fukugita et al.
1996; Stoughton et al. 2002) down to a limiting magnitude $g \approx
23$\,mag.

\subsection{Data reduction}
The {\it XMM-Newton} observations have been analysed using the Science
Analysis Software (SAS 5.3.3). Event files for the PN and the two
MOS detectors have been produced using  the {\sc epchain} and the
{\sc emchain} tasks of SAS respectively. These were then screened
for high particle background periods by rejecting times with
0.2-10\,keV count rates 
higher than 25 and 15\,cts/s for the PN and
the two MOS cameras respectively.  A total of 5 fields suffering
from  significantly elevated and flaring particle background were
excluded from the analysis. As a result our final {\it XMM-Newton}/2dF survey
comprises a total of 13 fields in the SGP and F864 regions
covering an effective area of $\sim \rm 2.3\,deg^{2}$. 
Note that only events  corresponding to patterns 0--4 for the PN 
and 0--12 for the two MOS cameras have been used.
The exposure maps have been created using the SAS task {\sc eexpmap}.
To increase the
signal--to--noise ratio and to reach fainter fluxes, the two MOS
event files have been combined into a single event list using the
{\sc merge} task of SAS.
We choose to analyse the images in the soft spectral band (0.3-2\,keV).
 This is because the groups and poor clusters have 
 low temperature and  hence their X-ray emission peaks at soft energies 
 ($<2$ keV). Therefore poor clusters  could remain undetected 
 in the total band images where the background is much higher.

\subsection{Source detection}
The source detection is performed on
the  MOS 1 and 2 merged image and the PN image
separately, by combining a wavelet detection algorithm and a maximum
likelihood fitting of the detected sources. We use the SAS {\sc
ewavelet} task\footnote{http://xmm.vilspa.esa.es/sas/current/doc/ewavelet/} with a
detection threshold of $5\sigma$ which detects both point and extended
sources  by convolving images with the Mexican hat filter (eg. Damiani
et al. 1997a, b). 
The scale of the Mexican hat filter used is 2, 4, 6, 8, up to 32 pixels.
 The {\sc ewavelet} task uses the exposure maps 
 to take into account sharp gradients in the exposure 
 (chip edges) and hence to avoid the detection of spurious sources.   
The output list is then fed into the SAS {\sc emldetect} task
which performs a maximum likelihood PSF fit on each source
yielding a likelihood for its extension. The
likelihood threshold corresponds to a probability smaller 
 than 0.001 of detecting 
spurious extended sources.
The detected sources in each field were then visually inspected and
detections clearly associated with CCD gaps, 
field of view edges or multiple point sources were removed. 

\section{The Extended Source Sample}\label{final}
The extended X-ray source sample,
compiled from the $5\sigma$ threshold 0.3-2\,keV source
catalogue, contains 8 objects with detected flux greater than 
$f_x(\rm 0.3 - 2 \,keV) \approx 2 \times10^{-14} \,erg \,s^{-1} \,cm^{-2}$.
In particular, we have (a) 3 SGP candidates in the merged MOS
mosaic and 3 in the PN (out of which all 3 overlap
with the MOS detections) and (b) 4 F864 candidates in 
the merged MOS mosaic and 4 in
the PN (out of which 3 overlap with the MOS detections).
In Table 1 we present the names, coordinates 
of the candidate X-ray clusters as well as the detector on
which the extended emission was identified. 
Visual inspection suggests that the 60\% overlap between the PN
and the MOS detected cluster candidates can be mostly attributed to 
the presence of gaps in the PN images and the elevated particle
background of this detector.
For example, candidate $\#$ 3, detected only on MOS, is located 
near one of the CCD gaps of the PN detector, which seems to be the reason for which it was not detected
in the PN image.

\begin{table*}
\begin{center}
\caption{Candidate cluster list in the XMM-Newton/2dF Cluster
  Survey, detected in either the merged MOS or the
  PN images. The correspondence of the columns is as follows:
index number, cluster name, celestial coordinates 
of the cluster center, the detector where the
  extended source was detected and corresponding names of already
  existing cluster lists.}\label{tbl1}
\tabcolsep 5pt
\begin{tabular}{cccc ccc}
\hline

$\#$ & Name & $\alpha$ & $\delta$ & Detector & Other Names\\
&  & (J2000)& (J2000)& & & \\ \hline 
1  & XMM2DF\,J134139.2+001739 & 13  41 39.2  &  +00 17 39 & MOS+PN & CE J205.412231+00.303271$^{1}$ \\
2  & XMM2DF\,J134304.8-000056 & 13  43 04.8  & --00 00 56 & MOS+PN & J1836.23\,TR$^{2}$  \\
3  & XMM2DF\,J134511.9-000953 & 13  45 11.9  & --00 09 53 & MOS & CE J206.296951-00.146028$^{1}$ \\
4  & XMM2DF\,J134413.8-002952 & 13  44 13.8  & --00 29 52 & MOS+PN & \\
5  & XMM2DF\,J134446.4-003019 & 13  44 46.4  & --00 30 19 & PN & \\
6 & XMM2DF\,J005656.8-274029 & 00  56 56.8  & --27 40 29 & MOS+PN & J1888.16\,CL$^{2}$  \\
7 & XMM2DF\,J005847.8-280027 & 00  58 47.8  & --28 00 27 & MOS+PN & \\
8 & XMM2DF\,J005623.2-281818 & 00  56 23.2  & --28 18 18 & MOS+PN & \\
\hline
\multicolumn{7}{l}{\footnotesize $^{1}$: Goto et al. (2002) SDSS; $^{2}$:
Couch et al. (1991)}\\

\end{tabular}
\end{center}
\end{table*}

Furthermore, note that had we accepted candidates with a lower
extension probability, the overlap between the detections in the merged MOS and
PN images would be significantly lower. This fact has guided us to
impose such a high probability limit for accepting candidates. The
 high limit although it secures that the resulting candidates are most probably real
X-ray clusters or groups, may also lead to  the omission X-ray faint or distant clusters. 
 Such is the case of cluster $\#$ 9 of Table 1
in Basilakos et al. (2004), which is also identified in the SDSS optical
data but has a lower extension probability (99\%) than the limit imposed in
the present work.

Out of these 8 cluster candidates, one ($\#$4 of Table 1) 
has been observed by {\it Chandra} and we have found that it consists
of three point sources blended, in the {\it XMM-Newton} image, in one 
 common extended envelope. 
This shows a possible problem from which our extended detections may
suffer. Therefore, source $\#$ 4 was excluded from our final
sample of X-ray cluster candidates.

We have performed a cross-correlation with all known cluster
catalogues using a search radius of 1.5 arcmin (corresponding to $\sim 0.5$
$h^{-1}$ Mpc at $z=0.4$) and we
have found that four out of our seven final cluster candidates have
been detected previously (either by Couch et al. 1991 or by Goto et
al. 2002). 

In Figure 1 we present optical images of our seven 
final X-ray clusters overlayed with their corresponding X-ray 
 contours. The contours were constructed from their
 smoothed X-ray image, using a Gaussian with a 1$\sigma$ radius
 of 3 pixels ($\sim 12.5$ arcsec). The four brightest clusters ($\#$ 1,
 2, 6 and 7) have successive contour width separation
 corresponding to 0.1 smoothed counts, while the fainter ones ($\#$ 3,
 5 and 8) have contour width separation of 0.05 smoothed counts.
 For the F864 northern clusters we use the optical $r$-band SDSS images, while
 for two of the SGP clusters we use the available DSS 
 optical images. For the cluster \#6 we use a WFI (Wide Field Imager)   
 r-band exposure (2400 s)  taken in 27  December 2000 at the Anglo-Australian
 Telescope. More details of these optical observations are given in
  Georgakakis et al. (2004).  

\begin{figure*}
\mbox{\epsfxsize=6.6in \epsffile{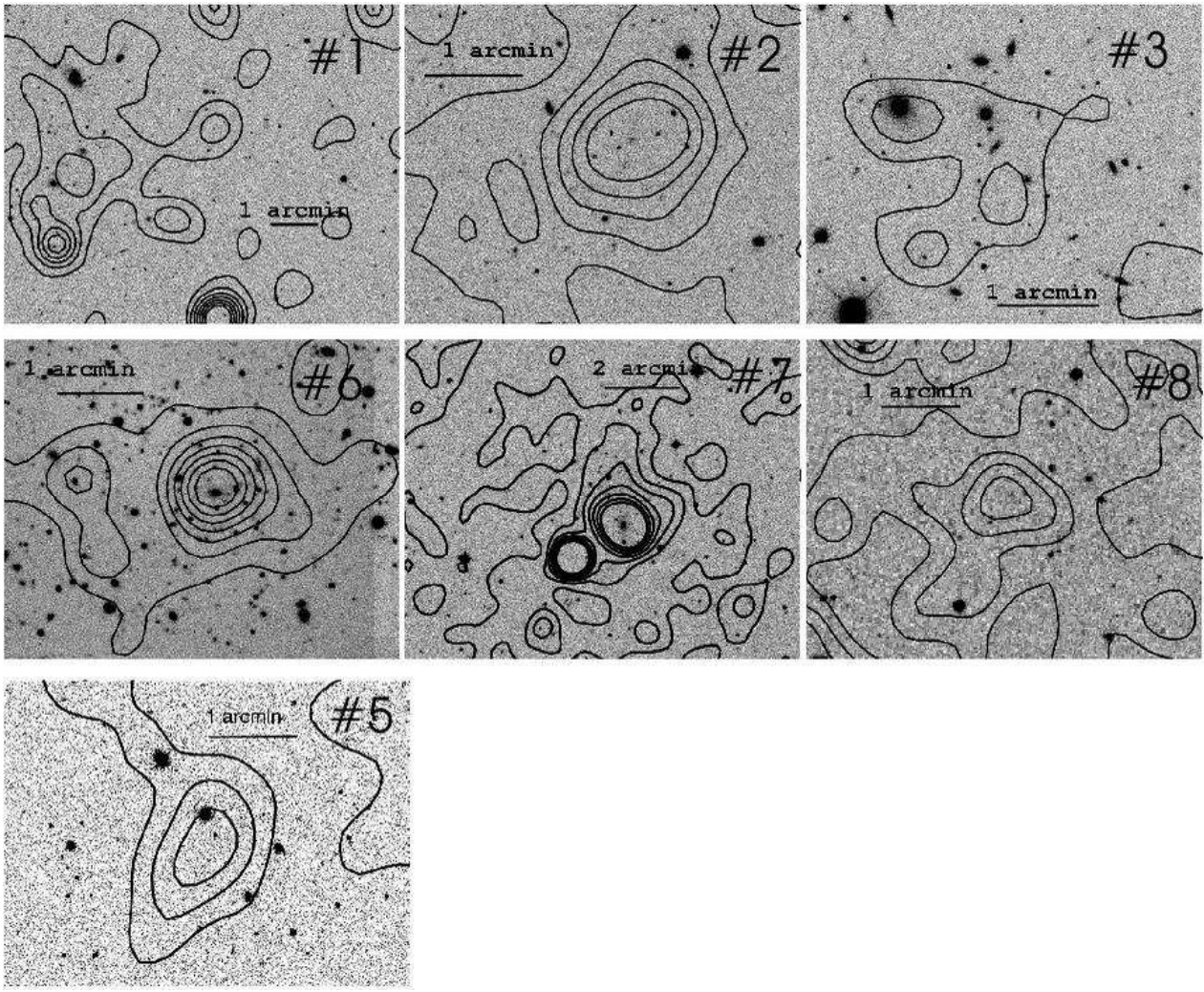}}
\caption{X-ray contours overlaid over optical images.
For the northern (F864) clusters ($\#$ 1, 2, 3), shown in the first row,
we use the $r$-band SDSS image. For the first cluster of the second
raw ($\#$ 6) we use a AAT/WFI r-band exposure while for the other two
($\#$ 7, 8) DSS Red images are used. Finally, in the third row 
we plot the northern cluster $\#$ 5 using the corresponding 
SDSS $r$-band image. The correspondnace between contours and smoothed
counts is indicated in the main text.}

\end{figure*}

\subsection{Flux Correction}
Since clusters of galaxies are extended X-ray sources,
their flux measured by any detection algorithm is only a
fraction of the total flux. To take into account the 
undetected low surface brightness emission in the far
wings of the source we apply a correction procedure 
which is based on fitting a King's profile (King 1962)
to the cluster X-ray surface brightness:
\begin{equation}
\sigma(r) = \sigma_{0}[1+(r/r_{\rm c})^2]^{-3\beta+1/2},
\end{equation}
where $\sigma(r)$ is the projected X-ray surface brightness as
a function of radius, $\sigma_{0}$ is the central X-ray surface
brightness, $r_{\rm c}$ is the core radius of the cluster and
$\beta$ is the ratio of the specific energy in galaxies to the
specific energy in the hot gas.
The true cluster integral source count rate, including the 
undetected flux, can be determined from:
\begin{equation}
\sigma_{true}(r) = 2\pi\int^{\infty}_{0}\sigma(r)r {\rm d}r =
\frac{\pi\sigma_{0}r_{\rm c}^2}{3(\beta-1/2)}
\end{equation}

Using the CIAO's {\sc Sherpa} task\footnote{http://cxc.harvard.edu/sherpa/} which
is a $\chi^2$ minimization procedure, we 
estimate the values of $\sigma_{0}$, $r_{\rm c}$ and the background
surface brightness.
Because of  the relatively few X-ray counts, we fix the value of ${\beta}$ 
to two possible values ($\beta=0.7$ and 1)
which are good approximations for relatively high
redshift clusters (eg. Arnaud et al 2002). 
We have verified that the convolution of the detector's PSF with the fitted
Kings-like profile does not significantly affect the resulting 
values of the core radii (difference $\le 4\%$).
For three of our
clusters which have either very irregular X-ray emission or very few
counts ($\#$ 1, 3 and 5) the King's-like profile fitting is not as
good, giving large $\chi_{\nu}^2$. values. However, the fitted values of their
core radii, $r_c$, are corroborated also by visual inspection 
of their X-ray extension. These are: $r_{c}\simeq 30^{''}$ ($\#$1), 
$r_{c}\simeq 10^{''}$ ($\#$3) 
and $r_{c}=\simeq 5^{''}$ ($\#$5).

We use the Energy Conversion Factors (ECF) of the individual detectors 
in order to convert count rates to flux. We assume 
a thermal emission ({\sc MEKAL}) spectrum with a 
temperature estimated from the X-ray spectrum 
(see section 3.2) and a Galactic absorption 
of $\rm N_H \approx 2 \times 10^{20} \rm
{cm^{-2}}$.

\subsection{X-ray spectral fitting}
We explore the X-ray spectral properties of our X-ray clusters using  the {\sc
xspec} v11.2 package to fit their  X-ray spectra.
The spectra from the three detectors (MOS1, MOS2 and PN)
were regrouped to have at least 25 counts per bin, thus ensuring 
that Gaussian statistics apply for 
the standard $\chi^{2}$ spectral fitting. 

\begin{figure*}

\rotatebox{270}{\epsfxsize=5.8cm \epsffile{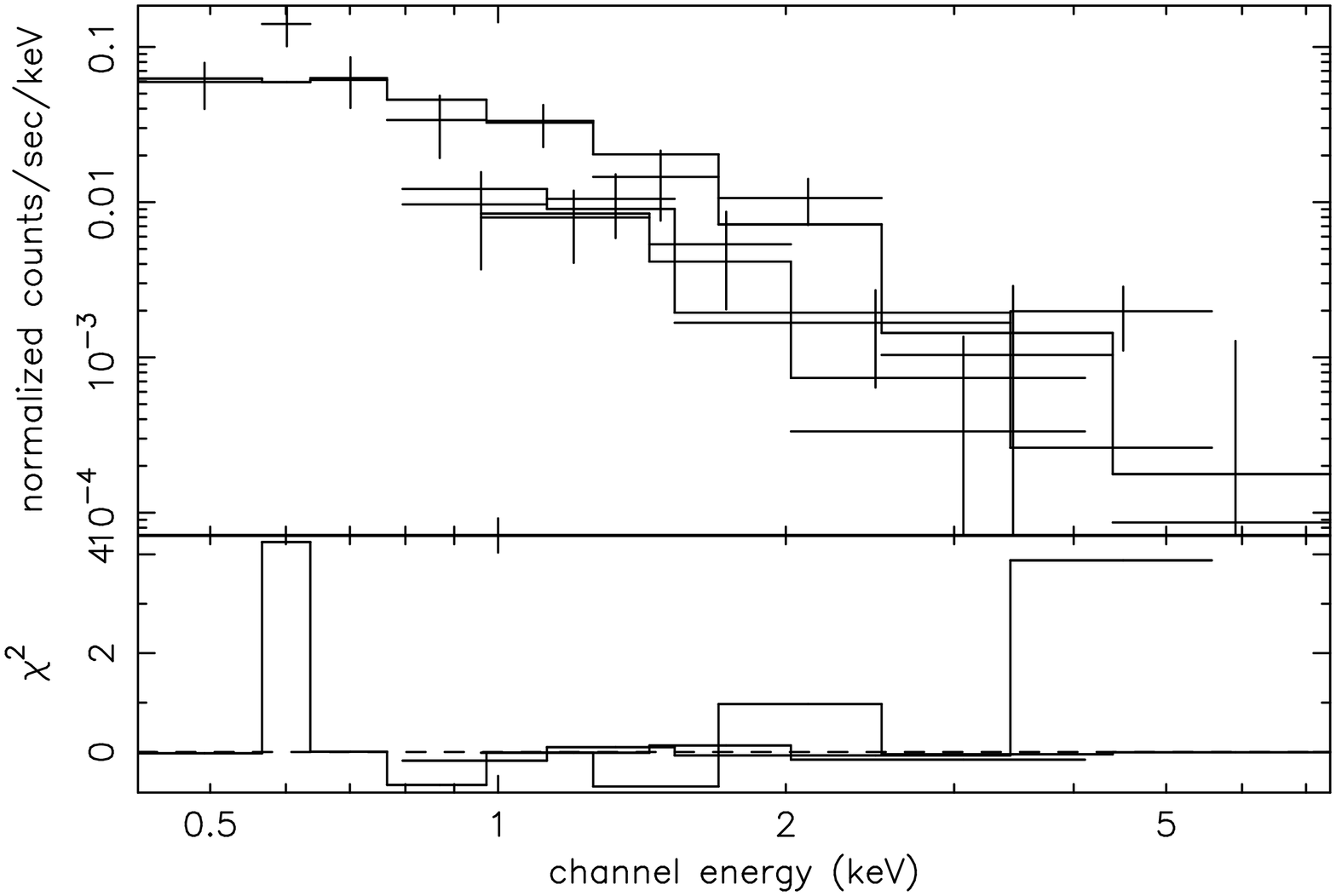}}
\rotatebox{270}{\epsfxsize=5.8cm \epsffile{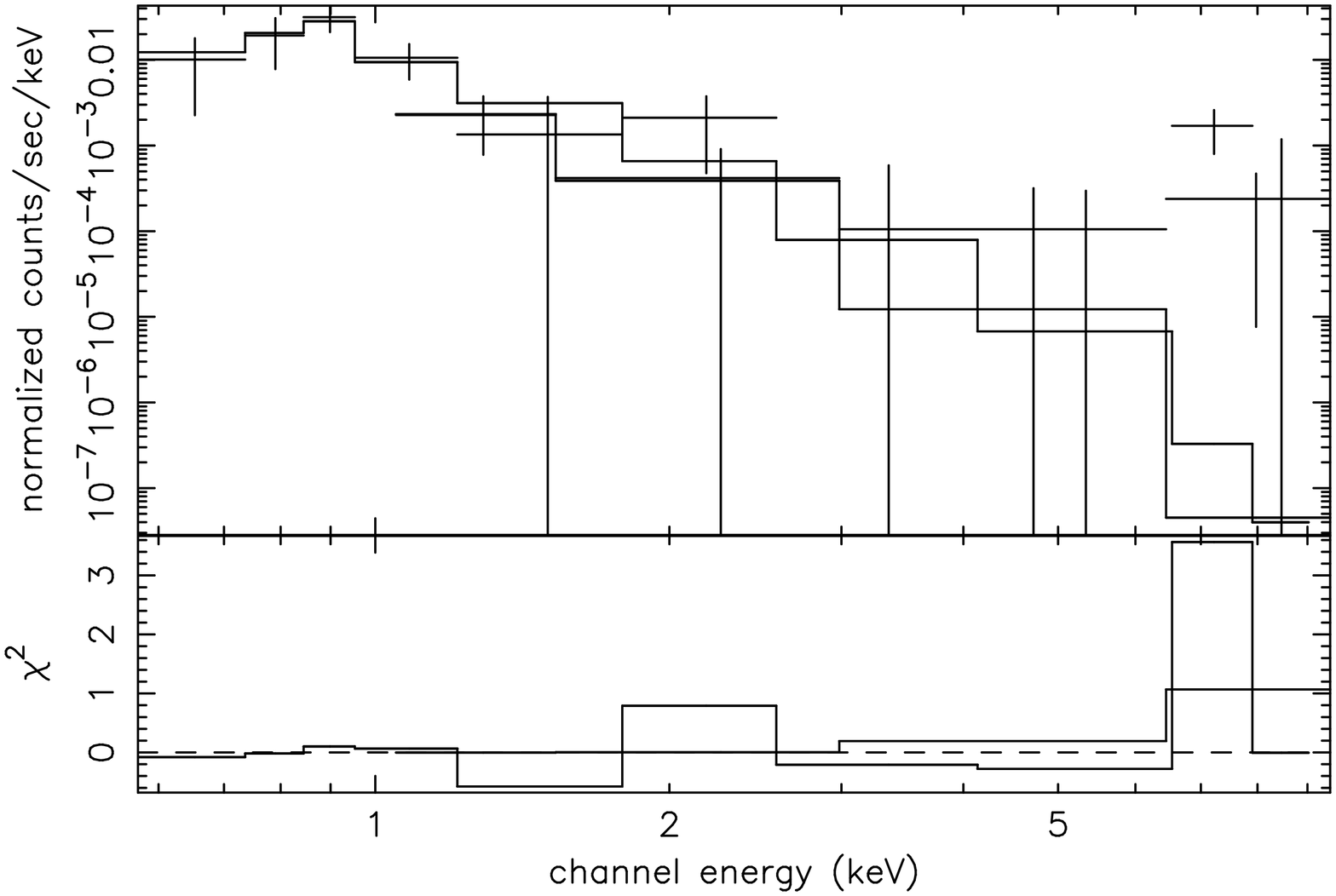}}
\vfill
\rotatebox{270}{\epsfxsize=5.8cm \epsffile{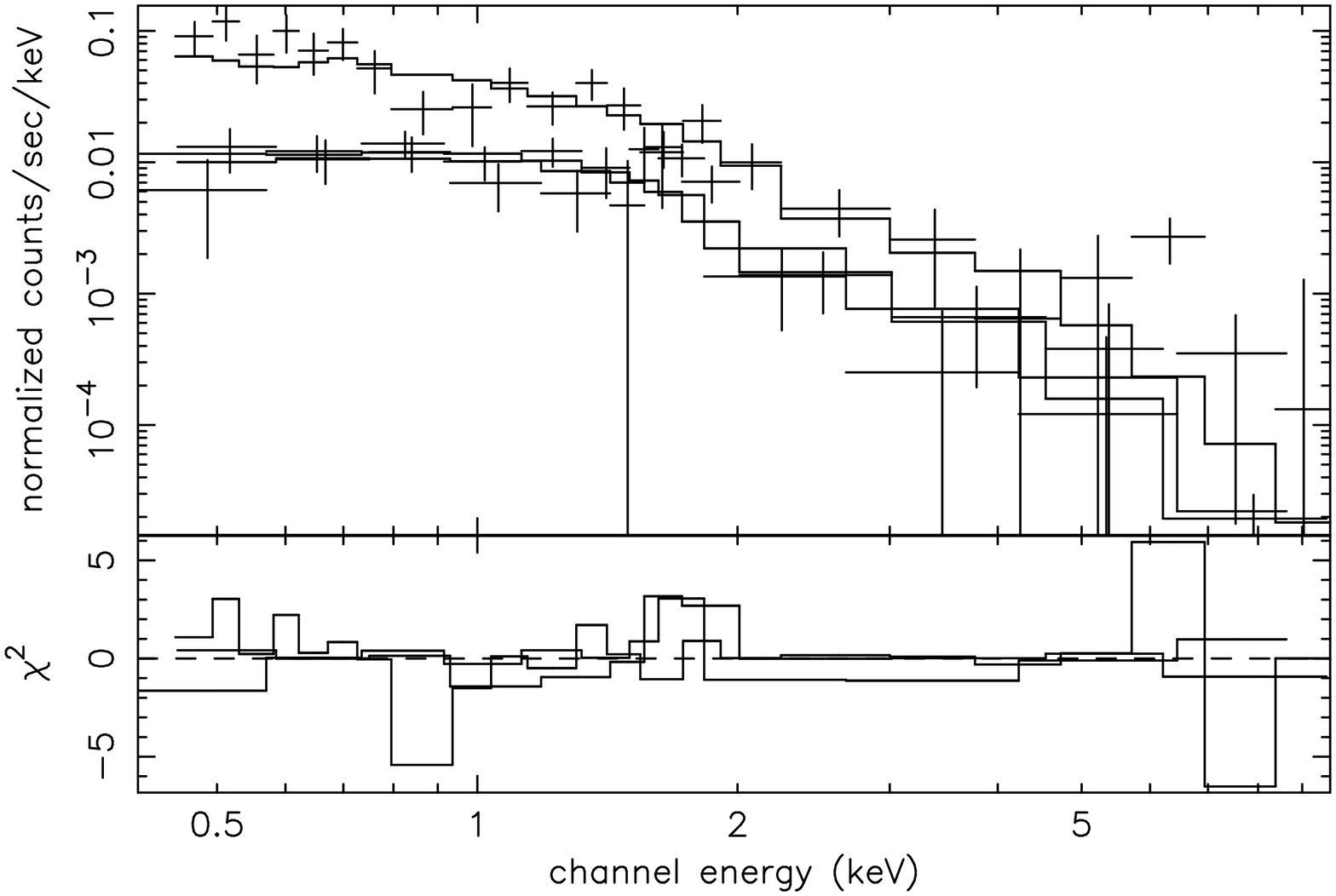}}
\rotatebox{270}{\epsfxsize=5.8cm \epsffile{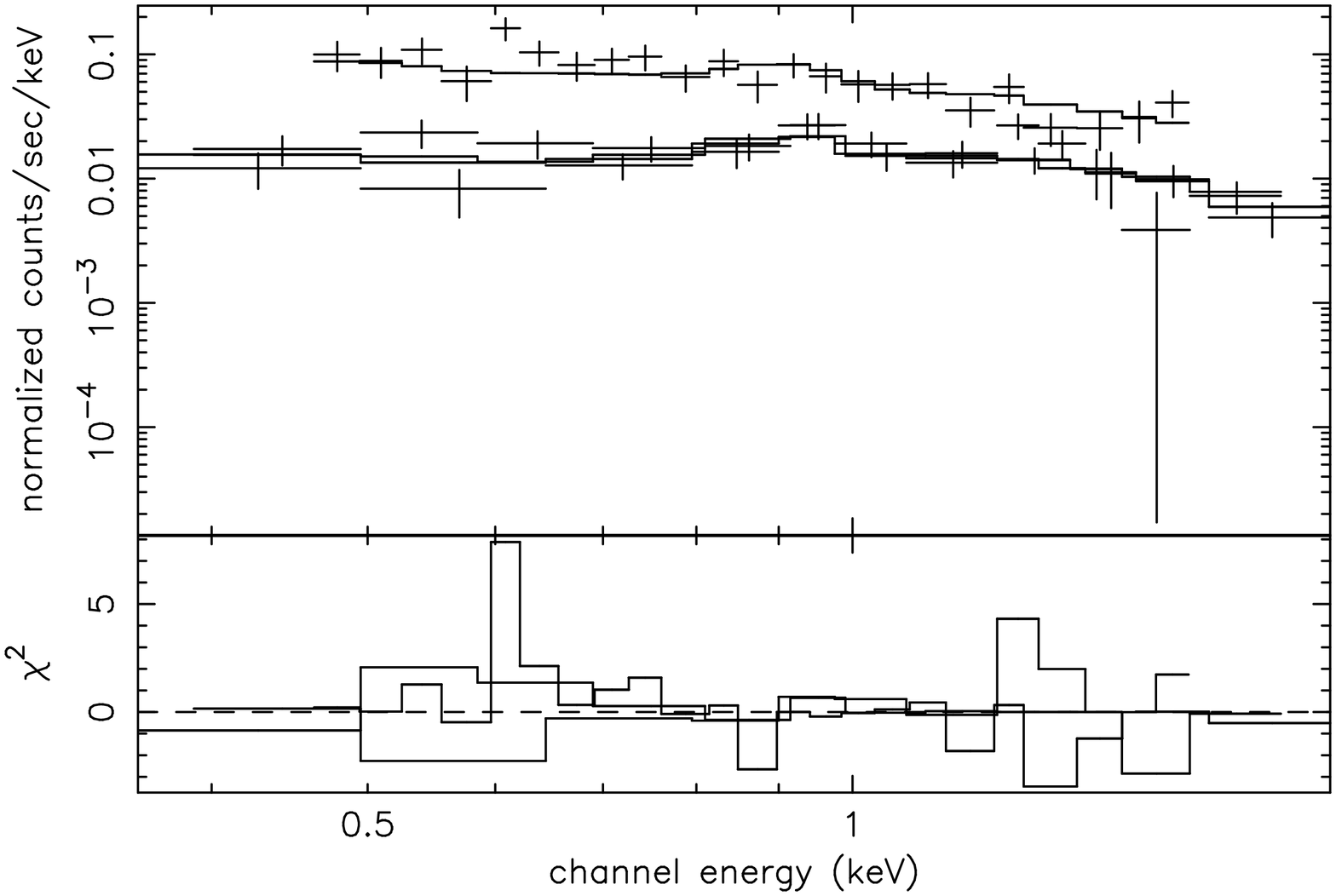}}

\caption{The X-ray spectra of clusters $\#$ 2 (upper left panel) and $\#$ 3
(upper right panel) and of clusters $\#$ 6 (lower left panel) 
and $\#$ 7  (lower right panel). The upper lines correspond to 
 the PN data while the lower lines to the MOS data. The $\chi^2$ 
 fitting residuals are plotted as well in the lower subpanels.}
\end{figure*}

For each cluster an X-ray spectrum was extracted from a region
large enough to include the cluster emission. A background
spectrum was taken from nearby source free regions. We removed in advance
the contribution of all point sources within the cluster and
the background regions. A photon redistribution matrix (RMF) and
ancillary region file (ARF) were created using
the {\sc SAS  rmfgen/arfgen} tasks.

To derive the global cluster X-ray characteristics, the binned
spectra in each instrument were fitted to a {{\sc MEKAL}) model of
thermal plasma emission with photo-electric absorption using XSPEC
(Arnaud 1996). We have
kept only two free parameters, the temperature and the
normalisation. The mean Galactic absorption of $N_H \simeq 2
\times 10^{20} \; {\rm cm}^{-2}$, 
the metal abundance of $Z=0.3 Z_{\odot}$ and the redshift were fixed. 

In Figure 2 we show the X-ray spectra for 4 of our 
cluster candidates.

\subsection{Individual Cluster properties}
The King's profile and the spectral fits were used to derive the cluster corrected flux,
luminosities and temperature 
and the results for the 7  cluster candidates are shown in Table
2. Note that the candidate X-ray clusters $\#$ 2, 6, 7 and 8 of Table
1 are well fitted by a King's
profile (with $P_{\chi^{2}}\ge 0.14$).
Also listed in Table 2 is the reduced $\chi^2$ of the spectral fit
while the errors presented are $2\sigma$ estimates.

\begin{table*}
\caption{ The final sample of the secure candidate X-ray clusters in the
{\it XMM-Newton}/2dF Cluster Survey. }\label{tbl2}
\tabcolsep 9pt
\begin{tabular}{lccccccccc}\hline
$\#$ & \multicolumn{2}{c}{$r_{\rm core}$ ($''$)} &
\multicolumn{2}{c}{$f_{x_{\infty}} \times 10^{-14}$} &
\multicolumn{2}{c}{$\log L_{x_{\infty}}$ (0.3-2 keV)}&  $kT$ & $\chi^2$/d.f. & $z$ \\ 
& $\beta=0.7$ & $\beta=1.0$ & $\beta=0.7$ & $\beta=1.0$ & $\beta=0.7$ & $\beta=1.0$ & (keV) & &\\  \hline 
1  & 30.0 & 30.0 & 19.4 $(\pm 1.9)$ & 15.4$(\pm 1.5)$ & 43.62 &43.52& 4.6$^{+8.7}_{-2.8}$& 1.46 & 0.39$^{1}$ \\
2  & 39.4 & 50.7 & 18.9 $(\pm 4.3)$ & 11.9$(\pm 2.7)$ & 44.12 & 43.92 & 2.6$^{+3.3}_{-1.2}$ & 0.93 & 0.68$^{2}$ \\
3  & 10.0  &  -  & 15.8 $(\pm 3.0)$ & -   & 42.16&  - & 1.0$^{+0.4}_{-0.3}$ & 0.63 &0.12$^1$ \\
5  & 15.0 & 15.0 & 2.5 $(\pm 0.8)$ & 2.0$(\pm 0.7)$ & - & - & - & - & -  \\
6  & 20.6 & 29.5 & 11.4 $(\pm 2.3)$ & 8.6$(\pm 1.0)$ & 43.73 & 43.61 &3.4$^{+1.4}_{-0.9}$ & 1.22 &0.56$^{2}$ \\
7  & 26.2 & 36.2 & 15.8 $(\pm 2.2)$ & 11.2$(\pm 1.5)$ & 42.90 & 42.75 &  2.6$^{+0.9}_{-0.4}$ & 1.10 & 0.20$^{3}$ \\
8  & 26.8 & 35.2 & 4.4 $(\pm 2.0)$ & 3.0$(\pm 1.4)$ & -  & - & -  & -  & -  \\ \hline
\end{tabular}

{\footnotesize $^{1}$ estimated by Goto et
  al. (2002); $^{2}$ spectroscopically confirmed by Couch et
  al. (1991); $^{3}$ {\sc 2df} redshift of central BCG with $m=18.7$, 
Colless et al (2001).}
\end{table*}

Note that XMM2DF\,J134139.2+001739 is also found in 
the SDSS optical data, by a variety of detection algorithms,
with an estimated redshift of $z\simeq 0.39$ by Goto et al. (2002) 
and $z\simeq 0.4$, based on the matched filter algorithm, by Basilakos
et al. (2004).
XMM2DF\,J134304.8-000056 was detected also
by Couch et al. (1991) and its redshift was spectroscopically measured
to be $z\simeq 0.68$ (detected on the SDSS data 
by Basilakos et al 2004 but not by Goto et al. 2002). 
XMM2DF\,J134511.9-000953 is most probably a
group of galaxies located at $z=0.12$, detected also by Goto et al. (2002)
The cluster J1888.16\,CL was identified also in {\it ROSAT}
data (Vikhlinin et al. 1998) while its spectroscopic redshift 
 was measured to be $z=0.56$ by Couch et al. (1991).

For XMM2DF\,J005847.8-280027 we provide a  prediction of the 
cluster redshift using the {\sc 2dF} redshift ($z\simeq 0.20$) 
of the BCG (called 2MASX J00584850-2800414), which is centered
on peak of the extended X-ray emission.
Finally, the XMM2DF\,J005623.2-281818 and
XMM2DF\,J134446.4-003019 
candidates have not been previously detected
and thus there is no reference in the literature of their redshift. The
relatively compact form of their X-ray emission together with the
faintness of the underline $r$-band galaxy distribution suggests
that they are very distant galaxy clusters, probably with $z\magcir 0.8$. 
Moreover, due to their low photon statistics we were not able to
estimate their temperature.

\subsection{The $\log N - \log S$ and $L_x-T$ relations}
We have attempted to construct the X-ray cluster $\log N - \log S$
relation and compare it with previous studies based on a {\it ROSAT}
deep cluster survey (Rosati et al. 1998). 
We estimate the survey sky coverage for our extended sources assuming 
a cluster size with a mean radius of 35 arcsec (8 pixels), 
which is typical of the extent of our extended sources. 
Sliding a circular aperture of that size across the 0.3-2\,keV background maps of
the survey area we estimate at each position
the $5\,\sigma$ background count fluctuations.
These are then divided with the
corresponding exposure time from the exposure map and converted to
flux adopting a bremsstrahlung SED with temperature $T\sim 3$\,keV and
Galactic absorption appropriate for each field. We finally correct
these fluxes for the emission outside the aperture used by 
adopting a King's surface brightness  profile with the individual
cluster core radii listed in Table 2. Note that 25 arcsecs correspond
to a core radius of $r_c\simeq 80 \; h^{-1}$ kpc at a characteristic 
depth of $z=0.4$.
The area in square degrees available to 
an extended cluster source of a given
0.3-2\,keV flux is shown in Figure 3. 
\begin{figure}
\mbox{\epsfxsize=3.in \epsffile{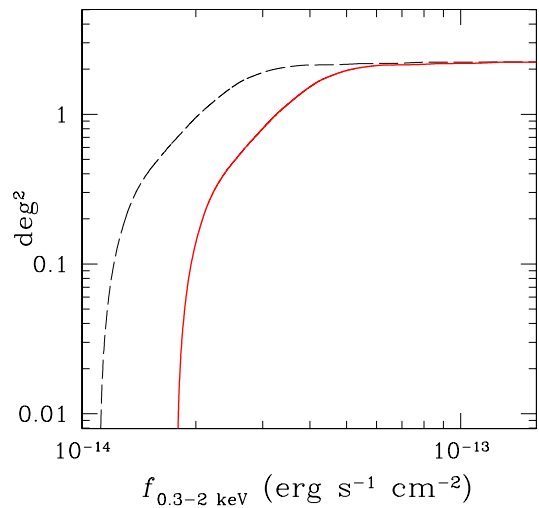}}
\caption{The area curves for the $\beta=0.7$ (solid line) and
  $\beta=1$ (dashed line) models, respectively.}
\end{figure}
The solid and dashed lines
correspond to models with $\beta=0.7$ and $\beta=1.0$ respectively,
also note that we have verified that the area curve is not
particularly sensitive to the choice of the aperture size used to sum
the background counts, the SED adopted to convert count rates to fluxes
or the correction factor to total flux. 

The X-ray cluster $\log N - \log S$  using the seven extended X-ray
sources in our sample is plotted in Figure 4, 
together with the Rosati et al. (1998) {\it ROSAT} 
based relation. In the left panel we plot the $\beta=0.7$ based
results while in the right panel the corresponding $\beta=1.0$
results. 
Within 1$\sigma$ there is a good agreement 
although our $\beta=0.7$ results appear to be somewhat higher
than those of Rosati et al. (1998). This is probably attributed 
to our small number statistics. 
\begin{figure}
\mbox{\epsfxsize=3.5in \epsffile{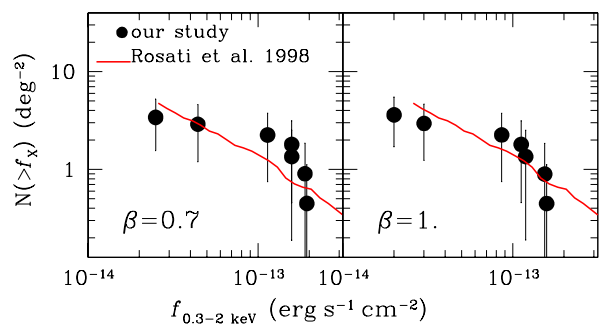}}
\caption{$\log N - \log S$ comparison of our survey (solid symbols) with
  the Rosati et al (1998) results (line). {\sc Left Panel}: Using a
  King's profile with $\beta=0.7$ and {\sc Right panel}: Using a
  King's profile with $\beta=1.0$.}\label{fig_lognlogs} 
\end{figure}

Finally, in Figure \ref{fig:Lx-T}
we plot the cluster $L_{x}$ against cluster $T_x$ (symbols) for
the $\beta=0.7$ case (see table 2).
Due to our poor statistics we do not perform a direct fit to
extract the $L_{x}-T$ relation. However, we show in Figure \ref{fig:Lx-T} 
the expected $L_x\propto T^3$ relation for $T>2$ keV with a normalization 
corresponding to the local $L_x-T$ relation (see Rosati et
al 2002) while for $T<1$ keV we plot the relation found by Helsdon \&
Ponman (2000). It is evident that our clusters roughly follow the expected
trends.

\section{Conclusions}\label{conclusions}
We have applied the SAS wavelet detection algorithm   
 on a wide field ($\sim  2.3 \, \rm deg^{2}$) shallow,
$f_X\rm (0.3 - 2\,keV)\approx 2 \times 10^{-14} \, erg \,
s^{-1}$, {\it XMM-Newton} survey to find extended emission
associated with clusters of galaxies.
After excluding spurious detections due to CCD gaps, image edges and
the blending of point sources we identified eight extended sources 
in either the merged MOS or PN images.  One of these
was excluded from our final list because using available {\it Chandra} data
it was found to be three point sources blended together.
 In the case where good quality optical observations are available, 
 galaxy overdensities were detected in 4 out of 5 cases. 
We have analysed the X-ray spectra of all seven candidates and derived 
an estimate of the cluster temperature for the five which had enough
photon statistics. We derive for the first
 time  the {\it XMM-Newton} cluster $\log N-\log S$; despite the 
 limited number statistics, and the $L_x-T$ relation which appear
to be in agreement with previous {\it ROSAT} results
(see Rosati et al. 2002). 

\begin{figure}
\mbox{\epsfxsize=3.in \epsffile{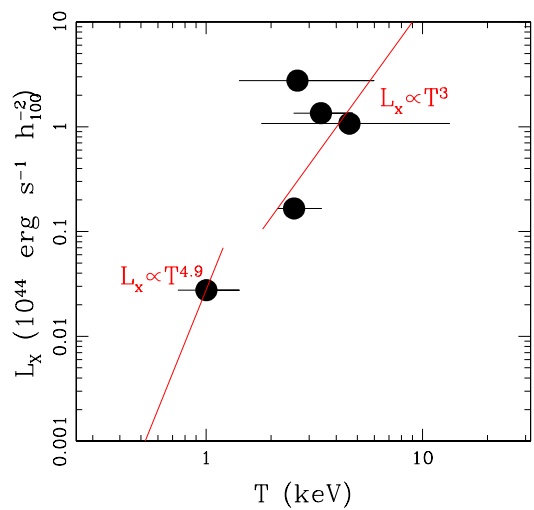}}
\caption{Our cluster bolometric $L_{x}$ - temperature relation (solid points) compared with
a fit to that of other surveys (see Rosati et al. 2002).}\label{fig:Lx-T}
\end{figure}

\section{Acknowledgments}
 This work is jointly funded by the European Union
 and the Greek Government  in the framework of the program
 ``Promotion of Excellence in Technological Development and Research'',
 project ``X-ray Astrophysics with ESA's mission XMM''.

\end{document}